*Essential upgrade of LOI 12-15-002 considered by PAC43 in 2015*

# Compton Edge probing basic physics at Jefferson Laboratory: light speed isotropy and Lorentz invariance

Vahe Gurzadyan[1,2], David Gaskell[3], Vanik Kakoyan[2], Cynthia Keppel[3], Amur Margaryan[2], Harutyun Khachatryan[2], Sergey Mirzoyan[2], Dipangkar Dutta[4], Branislav Vlahovic[2], Steve Wood[3]

1. NASA University Research Center for Aerospace Device, NCCU, Durham, NC, 27707, US
2. Alikhanian National Laboratory, Yerevan, 0036, Armenia
3. Thomas Jefferson National Accelerator Facility, Newport News, VA, 23606, US
4. Mississippi State University, P.O. Drawer 5167, MS, 39762, US

Spokespersons and contact persons: V. Gurzadyan, gurzadyan@yerphi.am,

D. Gaskell, gaskelld@jlab.org, A. Margaryan, amour@jlab.org

CONTENT

# Abstract

High precision testing of basic physical principles, among those the light speed isotropy and the Lorentz invariance have always been major goals of experimental projects, both of dedicated ones and performed within other specialized studies. The reason is natural, i.e. consequences of detections of possible violations or deviations from those principles will have a direct impact on the micro and macro physics, including cosmology and the nature of the mysterious dark energy. The development of Lorentz invariance violating (LIV) models has now become an established and active area of research, with periodical conferences and numerous publications, proceedings volumes and monographs. The profound observations in cosmology on the accelerated expansion of the Universe and the dark energy, dark matter and the cosmic microwave background tiny features, have made even more important the high accuracy testing of Lorentz invariance violating models. The recent discovery of the gravitational waves by the LIGO collaboration is also currently linked to the Lorentz violating models and the methods of its testing.

We propose to study of the light speed isotropy and Lorentz invariance at Jefferson Laboratory (JLab) by means of the measurements of the Compton Edge (CE) using of the Hall A/C existing experimental setup. Important motivation of this proposal is that, **methodologically the same experiment** (with participation of two co-authors of this proposal Gurzadyan, Margaryan) **has already been successfully elaborated at GRAAL** experiment at the European Synchrotron Radiation Facility (ESRF) in Grenoble with 6 GeV electron beam and the one-way light speed isotropy was tested to $10^{-14}$ accuracy. (Study of daily variations of CE at the Compton scattering of high energy electrons with monochromatic laser photons for this purpose was originally suggested in 1996 by Gurzadyan and Margaryan.) The limit for one-way speed of light isotropy and Lorentz invariance violation (LIV) coefficients obtained at GRAAL is currently a reference number for various extensions of Special Relativity.

This Proposal states two goals expected to be reached at JLab, both on Lorentz invariance: (a) the light speed isotropy testing accuracy, following from **conservative** evaluations at numerical simulations, about an order of magnitude better than was GRAAL's i.e. achieve few times $10^{-15}$; (b) the dependence of the light speed on the velocity of the apparatus (Kennedy-Thorndike measurement) will be traced to an accuracy $10^{-12}$, i.e. to about $10^3$ times better than available limits.

We outline the following issues regarding the aims in JLab vs ESRF-GRAAL:

(a) no principal methodological or technological issues can be expected for successful elaboration of the proposed program and obtaining of the sought results for higher energy electron beams in JLab, than it was in ESRF;

(b) A principal advantage in JLab as compared to ESRF is that, no temperature variations are expected in the tagging box, since in JLab one works with CW electron beam. That temperature effect was one of essential sources of systematic noise in ESRF, i.e. around 3 orders of magnitude larger than the scale of the sought effect, and needed much efforts to take it into account properly.

(c) the existing Hall A/C Compton (Møller) polarimeter setup will be used in essential way (from ESRF experience, a week measurements will solve the issue).

The performing of the CE studies with higher energy, i.e. up to 12 GeV beam in JLab due to the dependence on the square of the electron energy, will enable the increase the currently available values for the mentioned limits, with direct consequences on theoretical models in fundamental physics and cosmology.

This proposal (a) does not aim to review the existing LIV theoretical models and to convince in the importance of the LIV tests, which are widely discussed in quoted monographs, and conference proceeding volumes; (b) neither aims to provide a comparative analysis of methodologically different tests, since the matter is not in the competition of light speed isotropy or LIV tests and even of their accuracies but in their mutual complementation within the diversity of the physical phenomena involved.

We intend to proceed: (a) using initially the available experimental setups; (b) without requesting a dedicated beam time for the data storage as it was done initially at ESRF-GRAAL work.

In sum, this is a **low-cost** (parasitical, i.e. basically using the existing facility) and **feasibility confirmed** (at ESRF-GRAAL) Proposal with conservative estimation of one order of magnitude improvement for fundamental physics limitations.

Once green light is given to this Proposal, the full team membership (involvement of other experts), management (distribution of specific tasks, timetable, etc), asked funding estimations (too moderate), etc., can have sense to inquire into, again with the feasibility proof of ESRF.

1. Introduction

The continuous importance of probing of the accuracy of basic physical principles, the Lorentz invariance and the light speed isotropy, was stated by Einstein in 1927 [1], i.e. far after the classical Michelson-Morley (MM) experiment and the formulation of the Special Relativity. Since then, such studies were continuously in the agenda of various experiments, e.g. [2-7], parallel to the active theoretical studies of corresponding models, i.e. extensions of the standard model [8-11]; for extensive references both on the experiment and theory see http://en.wikipedia.org/wiki/Modern_searches_for_Lorentz_violation.

The indications of the dark energy in the Universe and the active search of the B-mode polarization of cosmic microwave background (CMB) have increased the interest to models with varying physical constants, including the speed of light, violation of the Lorentz, CPT invariance, e.g. [9-13].

The Compton Edge (CE) method for the light speed isotropy testing was suggested in [14] and includes the high precision daily measurements at the scattering of monochromatic accelerated electrons and laser photons to trace the one-way light speed. For comparison, majority of performed measurements of the light speed isotropy including the Michelson-Morley one, were dealing with a closed (two-way) path propagation of light (see [2-7], [11] and references therein). Such round-trip propagation is insensitive to the first order, but are sensitive only to the second order of the velocity of the reference frame of the device with respect to a hypothetical universal rest frame. Mossbauer-rotor experiments yield a one-way limit $\Delta c/c < 2 \; 10^{-8}$ [3], using fast beam laser spectroscopy. The latter using the light emitted by the atomic beam yields a limit $\Delta c/c < 3 \; 10^{-9}$ for the anisotropy of the one-way velocity of the light. Similar limit was obtained for the difference in speeds of the uplink and the downlink signals used in the NASA GP-A rocket experiment to test the gravitational redshift effect [2]. One-way measurement of the speed of light has been performed using also NASA's Deep Space Network [8]: the obtained limits yielded $\Delta c/c < 3.5 \; 10^{-7}$ and $\Delta c/c < 2 \; 10^{-8}$ for linear and quadratic dependencies, respectively. Another class of experiments dealt not with angular but the frequency dependence of the speed of light, as well the light speed dependence on the energy, using particularly the emission detected from the distant gamma sources, see the quoted Wikipedia article for extensive and recent references.

The CE method for this aim was first elaborated at the GRAAL facility in European Synchrotron Radiation Facility in Grenoble [15-18] using the electron beam of 6 GeV. First, the already stored data have been analyzed, then the decision on the move on a dedicated experiment has been drawn with the upgrading of the existing facility and monitoring. The final limit obtained based on the dedicated measurements of 2008 yields $\Delta c/c = 10^{-14}$. This result is currently a reference number to constrain various classes of theoretical models, thus proving both the experimental feasibility and profound motivation of the task.

Further lowering of that limit at the measurements at Jefferson National Laboratory for 11 GeV electron beams will enable to exclude certain models and pose constraints on the parameters of the others, with direct impact both on the models of dark energy, early evolution of the Universe and the interpretation of the observational/cosmological data regarding the fundamental physical principles.

Along with the repeating the GRAAL's CE measurements at JLAB we propose to probe also another principal effect, namely, the Kennedy-Thorndike experiment [19], [11], [20] on the light speed's sensitivity to the velocity of the apparatus, and linked to the Ives-Stillwell experiment of time delation [11]. This, as described below, is aimed to probe the possible the dependence of the CE on the electron beam energy using the Compton scattering as elementary process revealing momentum and energy conservation laws. The dependence of light speed $c(\vartheta,v)$ on the angle and the velocity of a moving frame v is given as [11]

$$c(\vartheta, v) = c\ [c_p(v) + c_t(v)]\ /\text{sqrt}\ [c_t^2(v) \sin^2(\vartheta) + c_p^2 \cos^2(\vartheta)], \qquad (1)$$

where $c_t$ and $c_p$ are the components of the light speed orthogonal and parallel to the velocity vector of the moving frame, respectively.

The study of the light speed anisotropy using the Compton Edge method with respect to the dipole of the cosmic microwave background radiation (CMB) as suggested in [14], as the modern analog of the Michelson-Morley experiment, is also linked with the determination of the hierarchy of inertial frames and their relative motions, and is defining an "absolute" inertial frame of rest, i.e. the one where the CMB dipole and quadrupole anisotropies vanish. Namely, the dipole anisotropy of the temperature *T* of CMB is of Doppler nature [22]

$$\frac{\delta T(\theta)}{T} = (v/c)\cos\theta + (v^2/2c^2)\cos 2\theta + O(v^3/c^3) \tag{2a}$$

(the first term in the right hand side is the dipole term) and is indicating the Earth's motion with velocity

$$v/c = 0.000122 \pm 0.00006; \quad v = 365 \pm 18 \, km/s, \tag{2b}$$

with respect to the above mentioned CMB frame. WMAP satellite [22] defines the amplitude of the dipole 3.346±0.017 mK and the coordinate of the apex of the motion (in Galactic coordinates)

$$l = 263.85 \pm 0.1; \quad b = 48.25 \pm 0.04. \tag{3}$$

This coordinate is in agreement with estimations based on the hierarchy of motions involving the Galaxy, the Local group and the Virgo supercluster [23].

The probing of the anisotropy of the speed of light with respect to the direction of CMB dipole, therefore, is a profound aim.

Now, let us outline the Compton Edge test in the context of the two-way and one-way tests of the light speed anisotropy and that of Lorentz invariance violation.

Einstein's Special Relativity posits the speed of light $c$ as:
a) exactly constant, strictly independent of the magnitude and direction of the velocity of the observer relative to any rest frame;
b) independent on the observation angle in the local frame.

These invariance features of $c$ are used to tie together the concepts of space and time, leading to the famous Lorentz transformations (Lorentz group). The Inverse Compton Backscattering (ICB) of laser photons on relativistic electrons has been considered as a sensitive test for tracing of the potential manifestations of Lorentz invariance violations [14]. Indeed, if the energy of an ultra-relativistic electron beam of Lorentz factor $\gamma$ is kept stable with given accuracy and over a reasonably long time and not at an instantaneous measurement, the Compton Edge $\omega_{21}^{max}$ ($E_{21}^{min}$) variation will result in the estimation of the light speed variation as follows from standard formulae (see [14], also below Sect.6 for details)

$$\beta \, d\beta = \left(\frac{1}{\gamma^2}\right) \frac{d\gamma}{\gamma} . \tag{33}$$

$$\frac{\delta c}{c} \sim \left(\frac{1}{\gamma^2}\right) \frac{\delta \omega_{21}^{max}}{\omega_{21}^{max}} \sim \left(\frac{1}{\gamma^2}\right) \frac{\delta E_{21}^{min}}{E_{21}^{min}}$$

Here the 'trick' is that, the error in c is reduced by a factor $(1/\gamma^2)$ with respect to the Compton Edge or beam energy error $(d\gamma/\gamma)$. The JLab CEBAF parameters, beam relative energy spread, $\delta E/E$, of a few $10^{-5}$ and a geometrical emittance of $10^{-9}$ m×rad for E = 11 GeV and $\delta E/E \sim 10^{-5}$, one has $\delta c/c \sim 2 \times 10^{-14}$. The error of mean values of $\delta c/c$ measured in this way should be improved by 1-2 orders of magnitude. Therefore, the kinematics of the Compton scattering of high energy electrons, i.e. of high Lorentz factor, and laser photons can be used for tracing of the potential manifestations of LIV.

The widely discussed experiments to test the STR may be divided into three types: (a) Michelson–Morley (M-M) which tests the isotropy of the speed of light, (b) Kennedy–Thorndike (K-T) which tests the velocity dependence of the speed of light, and (c) Ives and Stilwell (I-S), which tests the relativistic time dilation; see [11]. Most of these experiments, especially, those of M–M and K–T type test the two-way speed of light (in a closed path of a given length). However, still there are questions about the constancy of the one-way speed of light [24, 25]. Therefore, we now briefly compare the M-M (one-way and two-way), K-T, and I-S type measurements vs other LIV measurements. Let us represent the light speed variations as [20]

$$\delta c/c = \varepsilon_{KT} (v/c)^2 + \varepsilon_{MM} (v/c)^2 \sin^2\theta,$$

where $\varepsilon_{KT}$ and $\varepsilon_{MM}$ are the K-T and M-M contributions, respectively; obviously both vanish in Special Relativity. Recent ground experiments have obtained the result $\delta c/c \leq 10^{-15}$ and consequently $\varepsilon_{KT} = -4.8 \mp 3.7 \times 10^{-8}$ by comparing the frequencies of a cryogenic sapphire oscillator with hydrogen maser over a period of about 6 years [21]. A new space experiments are considered with improved technologies to improve this result by factor of $\sim 100$ (see [20] for more discussions). The current best result for M-M contribution $\varepsilon_{MM} \cong 10^{-10}$, e.g. [24].

The Compton scattering of laser photons with energies $\omega_{01}$ against high energy electrons with Lorentz factor $\gamma = (1 - \beta^2)^{-1/2}$ can be considered as a four step process:

1. In the Laboratory frame the laser photons with energies $\omega_{01}$ travel (see below Fig. 3, from right to left) with velocity $c_1$ (due to possible spatial anisotropy), electrons move with energies $E_1$ and velocity v;
2. In the rest frame of electrons the Doppler shifted (due to relativistic time dilation) laser photons with energies $\omega_{e1} = \gamma(1 - \beta\cos\theta)\omega_{01}$ scatter on electrons with velocity $c_2$ (due to potential velocity dependence), were $\beta = v/c_2$, $\gamma = (1 - \beta^2)^{-1/2}$. In our case, at head-on collisions $\theta = 0$ and $\omega_{e1} = \gamma(1 + \beta)\omega_{01}$;

3. In the rest frame of electron, 180° Compton scattered photons with energies $\omega_{ec} = \omega_{e1}/(1 + \frac{2\omega_{e1}}{mc^2})$, move away with velocity $c_3$ (because photons change their direction by 180°). The recoil electrons have equal and opposite momentum ($mc^2$ is the rest mass of electron);

4. Finally, in the laboratory frame we have the Compton scattered and Doppler shifted laser photons with energies $\omega_{21} = \gamma(1 + \beta\cos\theta_\gamma)\omega_{ec}$ and velocity $c_4$ (due to potential velocity dependence), where $\beta = v/c_4$, $\gamma = (1 - \beta^2)^{-1/2}$ and recoil electrons with energies $E_2 = E_1 - \omega_{21}$. In our case $\theta_\gamma = 0$ and for Compton Edges we have $\omega_{21}^{max} = \gamma(1 + \beta)\omega_{ec} = \gamma^2(1 + \beta)^2\omega_{01}/(1 + \frac{2\gamma(1+\beta)\omega_{01}}{mc^2})$ and $E_2^{min} = E_1 - \omega_{21}^{max}$.

In Special Relativity, obviously, $c_1 = c_2 = c_3 = c_4 = c$, where c is a universal physical constant of current adopted value 299 792 458 meters per second.

We propose to carry out two types of measurements at JLab by using the Moller Compton polarimeter. The first measurement is related to the one-way and two-way light speed isotropy, which will be checked by precise measurement of the relative changes of Compton Edge, as was done at GRAAL. All other possible sources of changes related to the Compton polarimeter or to the parameters of the electron and laser photon beams have to be taken under control.

The second measurement aims to verify the Lorentz transformation at high Lorentz factors ($\gamma \geq 2 \cdot 10^4$) to precision $10^{-3} - 10^{-4}$. This can be obtained by determination of the absolute values of the incident electron beam and the Compton Edge from Compton scattering of monochromatic laser photons with energies $\omega_{01}$ or relative values of the incident electron beam energy and the two Compton Edges from Compton scattering of two monochromatic laser photons with energies $\omega_{01}$ and $\omega_{02}$ with similar precisions [33]. This is achievable with the Moller Compton polarimeter and could improve the current limits of the K-T type LIV tests by $10^3 - 10^4$ times: then $\varepsilon_{KT}$ is expected to be evaluated to $10^{-11} - 10^{-12}$ accuracy; satellite test proposal [20] mentions as a goal the accuracy $10^{-10}$ (v there is the satellite's velocity).

Thus, the CE method allows M-M, K-T and I-S type high precision measurements. At JLab for M-M (one way and two-way) type experiment we expect to achieve $10^{-14}$-$10^{-15}$ for K-T and I-S measurements $10^{-11}$-$10^{-12}$ precisions, respectively.

The organization of this document is as follows. First, we review some of the classes of the Lorentz invariance violating models which have been directly affected by the GRAAL results. Then we outline the Compton Edge method, and the spectrometer. The GRAAL measurements are reviewed thereafter. The expected results and their discussion conclude the document.

## 2. Impact on fundamental theory and cosmology

Theoretical studies, including string theory and extensions of Special and General Relativities, predict violation of Lorentz invariance (see e.g. [8-11] and references therein). Various experimental studies, including using astronomical sources, have been conducted to probe the limits of those basic principles, since the increase in the accuracy of the available experimental limits will have a direct impact for excluding of particular theoretical models or constraining the parameters of the others.

Here, for illustration, we mention several theoretical studies which are crucially based on the GRAAL experimental results; far more references can be found in the quoted articles.

### 2.1 The Robertson-Mansouri-Sexl model

One of often discussed extensions Special Relativity is the model suggested by Robertson and Mansouri and Sexl (RMS), dealing with the generalization of the transformation from one inertial frame to another including the spatial anisotropy and angular dependence of the speed of light. In that model the one-way velocity of light as measured in the inertial frame in which the laboratory is at rest can be written as (see [11])

$$c(v,\theta) = c[1 - (1 + 2\alpha)(v/c)\cos\theta + O(v^3/c^3)], \qquad (4)$$

where α=1/2 for the Special Relativity.

Below is a Figure from the review [24] on RMS light speed isotropy and Lorentz violating model vs the increase of the accuracy of the experiments. The one-way experiments of the GRAAL type are particularly important for testing of this popular extension of the Special relativity models and estimating the so-called Robertson-Mansouri-Sexl coefficients.

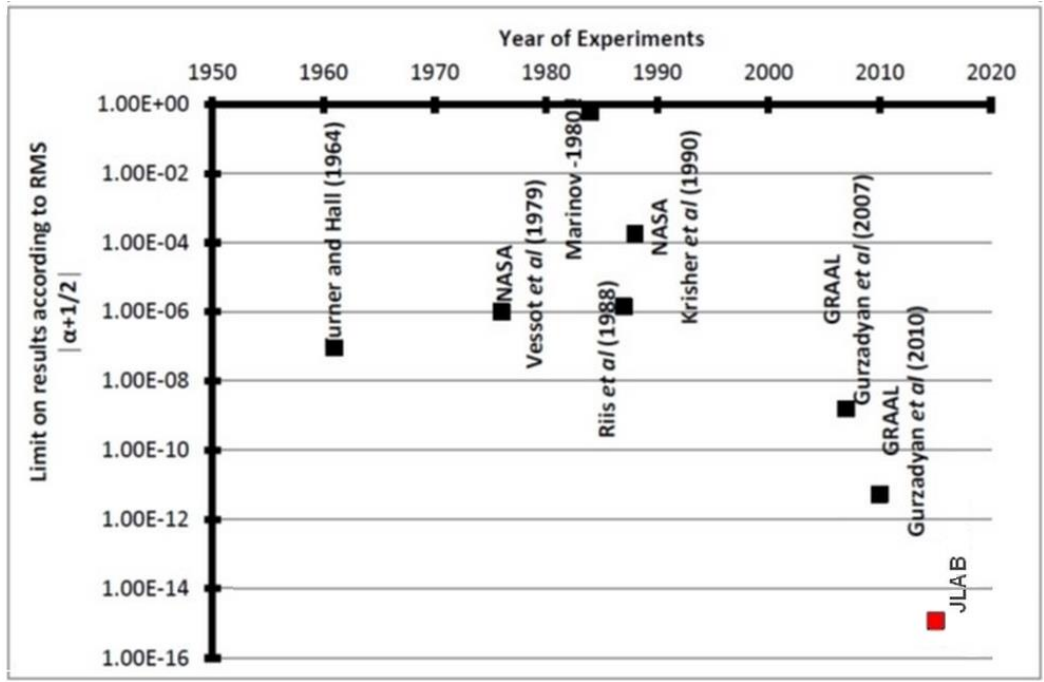

Fig. 1. This Figure reproduced from [24] exhibits the accuracy of the Robertson-Mansouri-Sexl coefficients obtained at various experiments. Einstein's Special theory of relativity corresponds to $|\alpha+1/2|= 0$. The red square is added here to denote the position of the expected result of JLab measurements as described in the current proposal.

**2.2 GRAAL results vs Standard Model Extensions (SME)**

In [25] the azimuthal dependence of the GRAAL data has been used for the diagnosis of a Standard Model Extension (SME) with space-time anisotropy, see the Fig. 2 below.

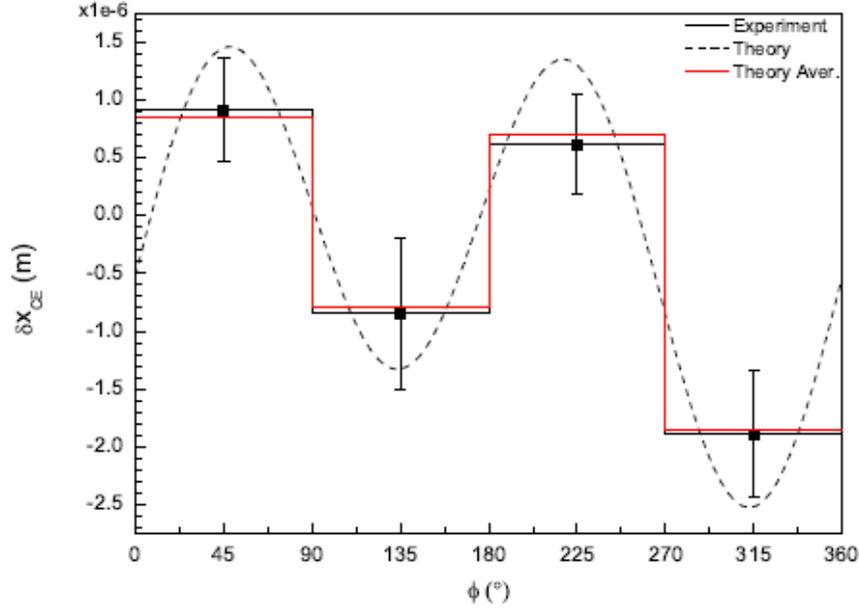

Fig. 2. δx$_{CE}$ azimuthal distribution vs angles of the GRAAL data of the year 2008 on a plane (*x-y* plane or θ=π/2), ξ = 3.64 × 10$^{-13}$, λ = 8.24 × 10$^{-14}$; this Figure is from [25].

On the one hand, the fit in the Figure 2 of the model with GRAAL data does not indicate much since the apparent variations in data are not statistically significant.

On the other hand, the unknown systematics in GRAAL experiments not allowing the further study of those apparent variations in the CE data can be overcome only at further increase of the accuracy of the measurements.

The principal conclusion is that, as one can see from Fig. 2, 4-fold increase in the accuracy, especially, by another facility, JLab, thus removing the instrumental systematics and will already be decisive for this model, either confirming the theoretically predicted variations or revealing their nature as of instrumental or other systematics.

**2.3. Minimal standard model extension: Finslerian photon sector**

The GRAAL results have been used also for obtaining of the constraints on the parameters of minimal standard model extension. Namely, in the minimal standard model extension (SME) the Lorentz-breaking Lagrangian of the pure photon sector is given by [26]

$$\mathcal{L}_{\text{photon}} = -\frac{1}{4}\eta^{\mu\rho}\eta^{\nu\sigma}F_{\mu\nu}F_{\rho\sigma} + \mathcal{L}_{\text{photon}}^{\text{CPT-even}} + \mathcal{L}_{\text{photon}}^{\text{CPT-odd}}, \tag{5}$$

where

$$\mathcal{L}_{\text{photon}}^{\text{CPT-even}} = -\frac{1}{4}(k_F)^{\mu\nu\rho\sigma}F_{\mu\nu}F_{\rho\sigma}, \tag{6}$$

$$\mathcal{L}_{\text{photon}}^{\text{CPT-odd}} = \frac{1}{2}(k_{AF})_\alpha \epsilon^{\alpha\beta\mu\nu}A_\beta A_{\mu\nu}. \tag{7}$$

The Lorentz invariance breaking parameter acquires an upper limit based on the GRAAL result

$$\sqrt{((\tilde{\kappa}_{o+})^{23})^2 + ((\tilde{\kappa}_{o+})^{31})^2} < 1.6 \times 10^{-14} \quad (95\% \text{ C.L.}) \tag{8}$$

### 2.4 Lorentz invariance violation with massive vector particles

Among the growing activity on Lorentz invariance violating models, which will be directly influenced by the proposed measurements, we mention the class of Standard-Model Extensions (SME) with massive vector particles violating Lorentz and CPT invariance, given by the Lagrangian [27]

$$\mathcal{L}_\gamma = -\frac{1}{4}F^2 - A \cdot j - \frac{1}{4}F_{\kappa\lambda}(\hat{k}_F)^{\kappa\lambda\mu\nu}F_{\mu\nu} + \frac{1}{2}\epsilon^{\kappa\lambda\mu\nu}A_\lambda(\hat{k}_{AF})_\kappa F_{\mu\nu}. \tag{9}$$

This extension is not only of interest for phenomenology of gauge bosons but also for Chern-Simons gravity emerging from string theory. The Chern-Simons gravity is among the models involved to explain the dark energy observational data.

### 2.5 B-mode cosmology and the break of Lorentz invariance

The detection of CMB's B-mode polarization will have major consequences not only for the studies of the early Universe, but for the fundamental physical principles up to Planck energies (cf. [11]). Among the theoretical models dealing with the scalar, tensor primordial fluctuations and the power spectrum of the temperature of the cosmic microwave background is the possible anti-correlation scalar and tensor modes via the Lorentz invariance violation [11]. The latter can be linked to large scale anomalies known in the temperature anisotropy defined by the two-point correlation function (for details and notations see [26])

$$C_l^{corr} = -\frac{\pi}{5}\frac{v^2}{\tau_0^2}\int \frac{dq}{k'^3}n^i n^j \prod_{ij}^{33}(k') = -\frac{\pi v^2}{75\, l^4}\left(-l_x^2 + 5l_y^2 + 2(3l_x^2 + l_y^2)\cos 2\theta\right). \tag{10}$$

The Lorentz invariance here is reduced to the invariance of this correlator with respect to rotations, i.e. which in principle can be tested at cosmic microwave background temperature measurements. Any constraint on the Lorentz invariance violation from the CE measurements can be confronted with these data and the interpretations both in the cosmological evolution context and basic physical principles up to Planckian energies.

The mentioned examples do indicate that, the increase of accuracy of measurements performed in JLab, in certain cases already 4-fold, can be crucial either to rule out certain theoretical models or to reveal the nature of the angular-dependent variations in the GRAAL data, with direct consequences for quantum field theories and cosmology.

### 2.5 Lorentz violating dark matter and dark energy

Dark energy and dark matter inducing Lorentz invariance violation models have also been investigated, see e.g. [28, 29] and references therein. In the model considered in [28] the low energy action for the scalar field has the form

$$S_{[\Theta]} = \int d^4 x \sqrt{-g}\left(-\frac{g^{\mu\nu}\partial_\mu\Theta\partial_\nu\Theta}{2} + \kappa\frac{(u^\mu\partial_\mu\Theta)^2}{2} - \mu^2 u^\mu \partial_\mu \Theta\right) \tag{11}$$

with the cosmological constant setting zero, and with the coupled field and the gravity of Einstein-Hilbert Lagrangian. Then the Friedmann equation has the form

$$H^2 = \frac{8\pi G_{cos}}{3}\left(\rho_\mu + \rho_s + \rho_d + \sum_{other}\rho_n\right), \tag{12}$$

where

$$\rho_\mu \equiv \frac{\mu^4}{2(1+\kappa)}, \quad \rho_s \equiv \frac{C^2(1+\kappa)}{2a^6}, \quad \rho_d = -\frac{\mu^2 C}{a^3}, \tag{13}$$

and the density components under the sign of the sum denote the standard matter components (cold dark matter, photons, neutrinos). The rest refer to those of the scalar field described by the dynamical equation

$$\dot{\Theta} = -\frac{\mu^2 a}{1+\kappa} + \frac{C}{a^2},  \qquad (14)$$

where C is the integration constant (for details see [28]). It is shown that the Lorentz violation, triggering a preferred direction in space-time, i.e. generating anisotropic stress, can be pronounced also in the evolution of cosmological perturbations, i.e. in the large scale effects.

In the case of dark matter the coupling part of the action has the form [28]

$$S_{[DMu]} = -m \int d^4x \sqrt{-g}\, n\, F(u_\mu v^\mu),  \qquad (15)$$

where *m* is the mass of the dark matter particles, *n* is the their number density and *v* is their 4-velocity, and the function *F* is entering the dynamical equations.. The extension of the special relativistic relation then reads

$$E^2 = m^2 c^4 + (1+\xi) p^2,  \qquad (16)$$

where ξ=0 corresponds to the Lorentz invariance.

The common in these models is the spatial anisotropy due to the Lorentz invariance violation determined by the scalar field which can reveal itself in cosmological scale as either dark energy or dark matter, or other observable effects such as the cosmic microwave background, the large scale matter distribution features, see also [29, 30]. It is noted that the latter are solely due to the scalar perturbations while the vector ones can lead to B-mode effects.

These are examples of models of link of the Lorentz invariance violation and the dark sector – dark energy and dark matter – properties in the cosmological scales. These models do involve options in the Lorentz violation schemes, each depending on defined parameters which themselves in each case at further specification can be constrained by the observational data on the cosmic background radiation, and the GRAAL type data or of future JLAB ones. No doubt, other specific models based on Lorentz invariance violation will be developed in future as well, and therefore its experimental limits will in either case affect the understanding of the dark sector.

3  **The Compton scattering of laser photons on the high energy electron beam**

The energy of the scattered photon $\omega_{21}$ is related to the energy of the primary laser photon $\omega_{01}$ by the equation (see [14])

$$\omega_{21} = \frac{(1-\beta\cos\theta)\omega_{01}}{1-\beta\cos\theta_\gamma + (1-\cos\theta_0)(\omega_{01}/E_1)}, \qquad (17)$$

where $E_1$ the energy of incident electrons, $\beta = v/c$, $v$ is the velocity of the incident electron, $\theta_0$ is the angle between the momentum of incident and scattered photon, $\theta$ and $\theta_\gamma$ are angles between the momentum of the electron and the incident and the scattered photons, respectively, $\theta_e$ is the angle between the momentum of incident and scattered electron (see Fig. 3). Rewrite this expression in the following form

$$\omega_{21} = \omega_{01} A \gamma^2, \qquad (18)$$

$$\gamma^2 = (1-\beta^2)^{-1} \qquad (19)$$

$$A = \frac{(1-\beta\cos\theta)(1+\beta\cos\theta_\gamma)}{1+\beta^2\gamma^2\sin^2(\theta_\gamma) + \gamma^2(1-\cos\theta_0)(1+\beta\cos\theta_\gamma)(\omega_{01}/E_1)}. \qquad (20)$$

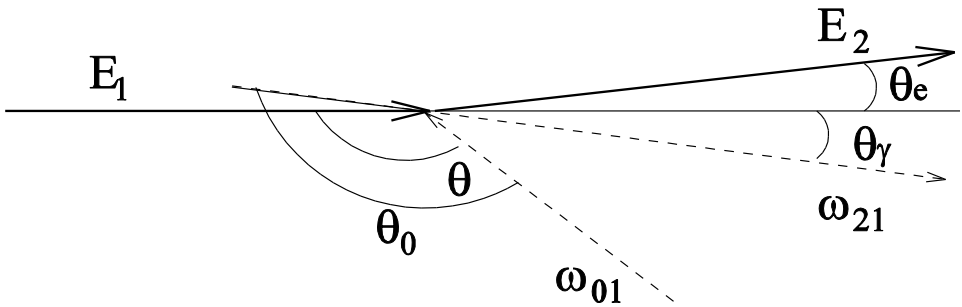

Fig. 3. The kinematics of Compton scattering.

The energy of the scattered photon at small angles $(\gamma\theta_\gamma) < 1$ up to $\gamma \sim 10^5$ is proportional to the square of the electron energy since the factor A depends on $\gamma \leq 10^5$ weakly.

Rewrite equation (18) in the following form [32, 33]

$$\omega_{21} = \frac{\omega_{01}}{1+(\theta_\gamma/\vartheta_0)^2} , \qquad (21)$$

$$\vartheta_0 = \frac{mc^2}{E_1}\sqrt{x+1} , \qquad (22)$$

$$X = \frac{4E_1\omega_{01}\cos^2(\alpha/2)}{m^2c^4}, \qquad (23)$$

where $\alpha = \pi - \theta$, m is the mass of electron.

The maximum (minimum) energy of the scattered photon (electron) or the Compton Edges of photons (electrons) are given by

$$\omega_{21}^{max} = \frac{x}{x+1}E_1 \qquad (24)$$

$$E_{21}^{min} = E_1 - \frac{x}{x+1}E_1 = \frac{1}{x+1}E_1 \qquad (25)$$

and

$$X = \frac{4E_1\omega_{01}}{m^2c^4} \qquad (26)$$

which is obained for 180° scattering in head-on collision

$\theta = \theta_0 - \pi$ and $\theta_\gamma = 0$.

For example: $E_1 = 6.0\ (11.0)$GeV, $\omega_{01} = 3.54\ (2.33)$eV, $x = 0.325\ (0.393)$, $\omega_{21}^{max} = 0.245\ (0.282)E_1$ and $E_{21}^{min} = 0.755(0.718)E_1$.

The photon and electron scattering angles are functions of the photon energy

$$\theta_\gamma(y_\gamma) = \vartheta_0\sqrt{\frac{y_\gamma^{max}}{y_\gamma} - 1}, \qquad (27)$$

$$\theta_e(y_\gamma) = \theta_\gamma\frac{y_\gamma}{1-y_\gamma}, \qquad (28)$$

where $y_\gamma = \omega_{21}/E_1$. These functions for $x = 0.325\ (0.393)$ are displayed in Fig. 4.

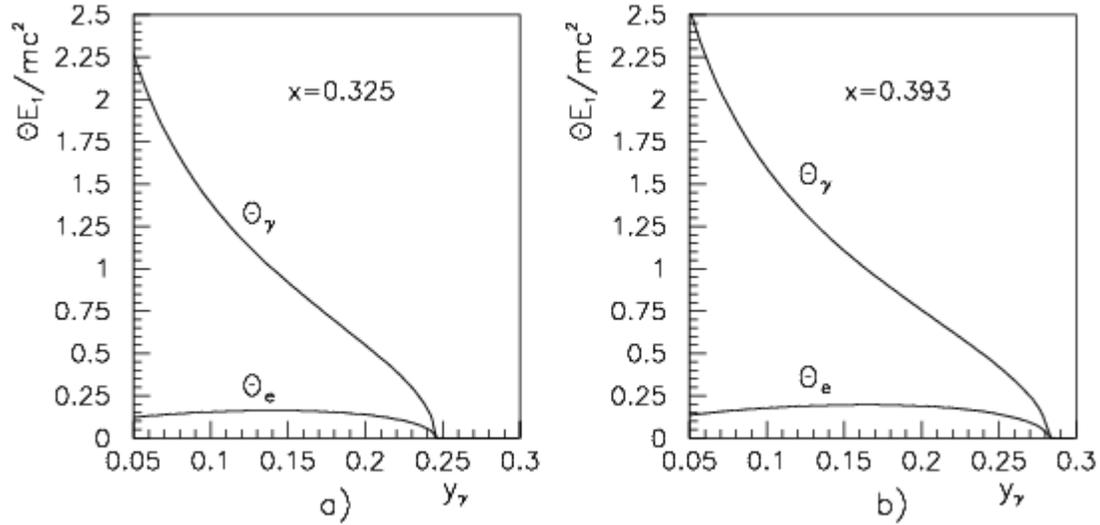

Fig. 4. Electron and photon scattering angles vs photon energy for: a) $E_1 = 6.0$ GeV, $\omega_{01} = 3.54$ eV and b) $E_1 = 11.0$ GeV, $\omega_{01} = 2.33$ eV.

The energy spectrum of the scattered photons is defined by the cross section

$$\frac{1}{\sigma_c}\frac{d\sigma_c}{dy_\gamma} \equiv f(x, y_\gamma) = \frac{2\sigma_0}{x\sigma_c}\left(\frac{1}{1-y_\gamma} + 1 - y_\gamma - 4r(1-r)\right), \tag{29}$$

where

$$y_\gamma \leq y_\gamma^{max} = \frac{x}{x+1}; \tag{30}$$

$$r = \frac{y_\gamma}{x(1-y_\gamma)} \leq 1; \tag{31}$$

$$\sigma_0 = \pi\left(\frac{e^2}{mc^2}\right)^2 = 2.5 \times 10^{-25} \text{cm}^2. \tag{32}$$

The total Compton cross section for the nonpolarized case is

$$\sigma_c = \frac{2\sigma_0}{x}\left(\left(1 - \frac{4}{x} - \frac{8}{x^2}\right)\ln(x+1) + \frac{1}{2} + \frac{8}{x} - \frac{1}{2(x+1)^2}\right). \tag{33}$$

The energy spectra of the scattered photons and electrons for $x = 0.325$ and $x = 0.393$ are displayed in Fig. 5 and Fig. 6, respectively.

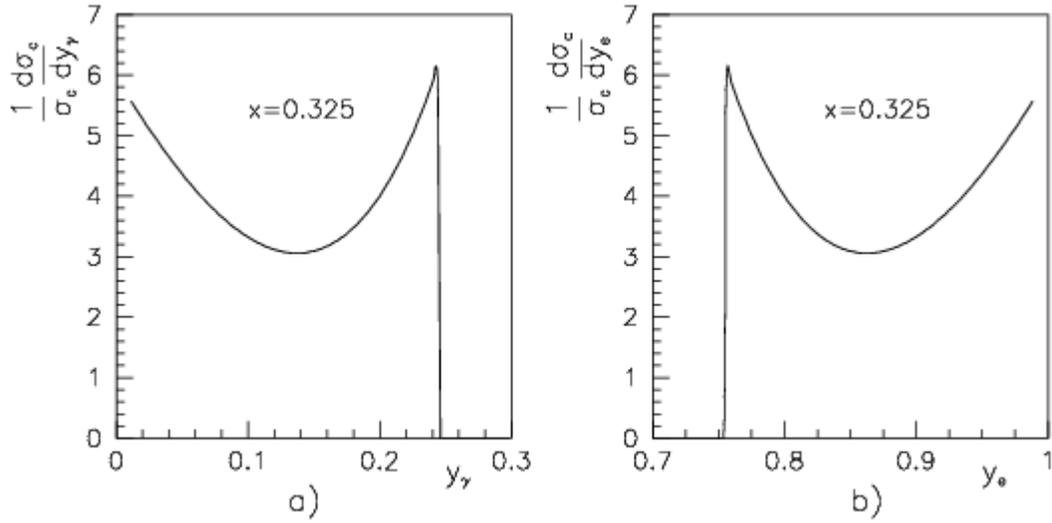

Fig. 5. Energy spectra of scattered photons and electrons for $E_1 = 6.0$ GeV, $\omega_{01} = 3.54$ eV.

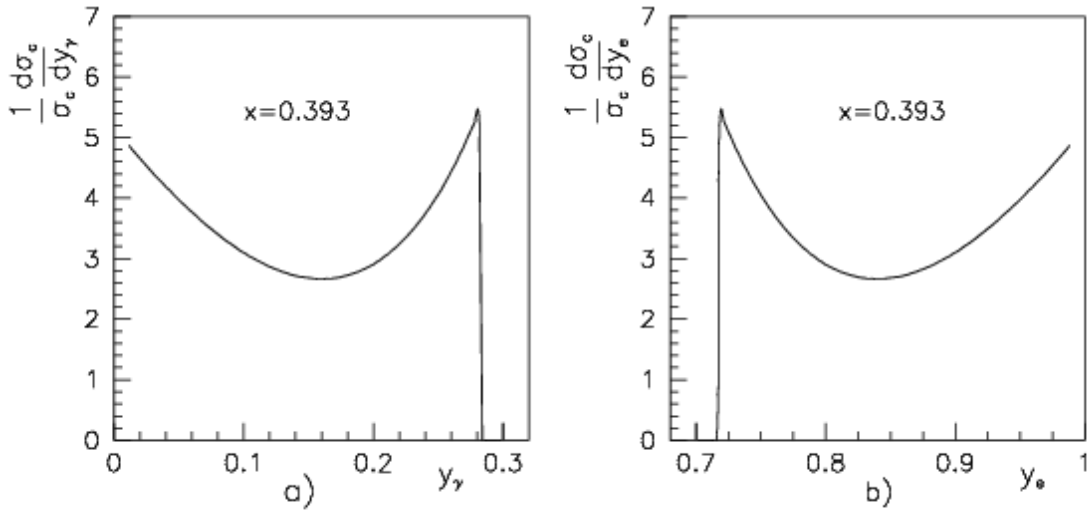

Fig. 6. Energy spectra of scattered photons and electrons for $E_1 = 11.0$ GeV, $\omega_{01} = 2.33$ eV.

The JLab CEBAF [34, 35] has the most advantageous parameters for the proposed experiment, since its beam has a relative energy spread, $\delta E/E$, of a few $10^{-5}$ and a geometrical emittence of $10^{-9}$ m × rad. For E = 11 GeV and $\delta E/E \sim 10^{-5}$, one has $\delta c/c \sim 2 \times 10^{-14}$. The error of mean values of $\delta c/c$ measured in this way should be improved by 1-2 orders of magnitude.

## 4 Compton Edge Electron Beam Energy Spectrometer

The main idea of this method based on the inclusive energy spectrum measurement near the kinematical (Compton) edge of electrons scattered on the laser photons. In this section we present the results of Monte Carlo simulations of such spectrometer. The energy spectrum of Laser Compton Scattered γ-rays and electrons were simulated with the Monte Carlo code based on the two-body kinematics by using the EGS4 [36] code and taking into account the energy spread of scattered electrons. The energy spread of LCS electrons includes the energy resolution of the recoil electron spectrometer and all the effects due to the electron beam phase space.

MC spectra of the LCS electrons of the GRAAL ($E_1 = 6000.0\ MeV, \sigma_E = 6.4\ MeV$) and JLab ($E_1 = 11000.0\ MeV, \sigma_E = 35\ MeV$) are displayed in Fig. 7a and Fig. 8a, respectively. Total number of events in these simulations is $5 \times 10^6$. Two fitting algorithms were used to extract the Compton Edges, of LCS electrons, i.e. the Lorentz factor of the incident electron beam. In the our previous simulations [33], the Lorentz factor of the incident electron beam were extracted as a single fitting parameter, fitting the MC spectrum by the convolution of the theoretical cross section with the resolution function Eq. (33). By this way it was demonstrated that the energies of the electron beams electron can be determined with a precision $10^{-4}$ or better [33, 35]. This result is close to the precision of Compton Edge obtained at GRAAL experiment (see Section 6). However to extract the Compton Edge at GRAAL experiment, a different fitting algorithm was used [37]. It consists in defining following figures $Y_i = N_{i+1} - N_i$ between two consecutive bins from the energy distribution of scattered electrons near Compton edge. Distributions of the $Y_i$ values, for the cases of the GRAAL and JLab, simulated from MC results presented in Fig. 7a and Fig. 8a, are displayed in Fig. 7b and Fig. 8b, respectively. Fitting these spectrums by the Gaussian function $F(x) = a\,exp(-0.5\left(\frac{x-b}{\sigma}\right)^2)$, the parameters related to the number of events (a), position of the Compton edge (b), and the effective energy resolution of the electron spectrometer (σ) can be extracted. The effective energy resolution included the energy resolution of the electron spectrometer and the energy spread of the Compton scattered electrons due to energy and angular spread of the incident electron and photon beams. From the distributions presented in Fig. 7b and Fig. 8b, for the extracted Compton edges we have: $E_{21}^{min} = 4527.0 \mp 0.13$ (7898.0 ∓ 0.4) MeV at expected theoretical values $E_{21}^{min} = 4527.05$ (7898.80) MeV for GRAAL (JLab) experiments.

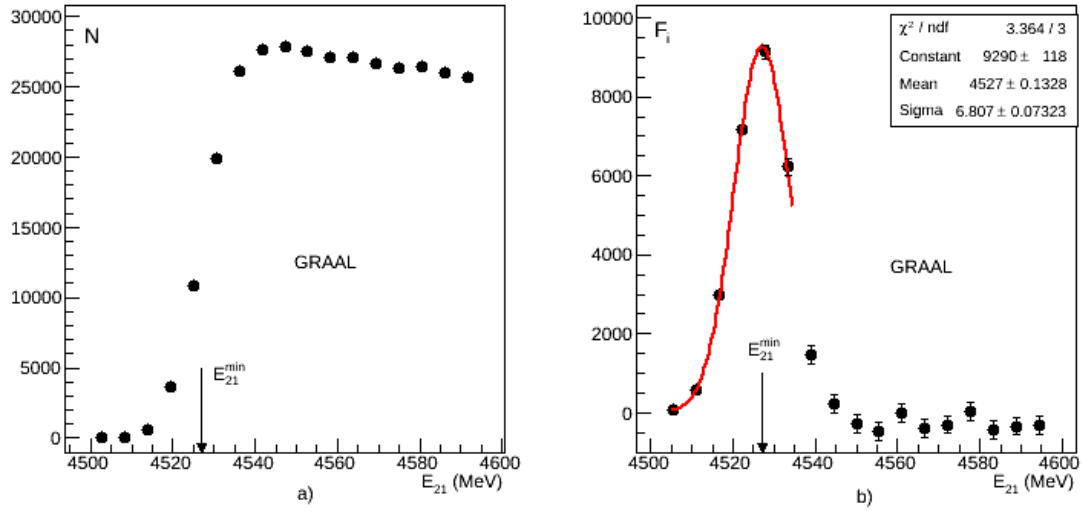

Fig. 7. a) MC energy spectrum of scattered electrons near Compton edge for GRAAL experiment: $E_1 = 6000.0$ MeV, $\omega_{01} = 3.54$ eV. Total number of events is $5 \times 10^6$. b) Distribution of the $Y_i = N_{i+1} - N_i$ values. The extracted electron Compton edge is $E_{21}^{min} = 4527 \mp 0.13$ MeV with an expected theoretical $E_{21}^{min} = 4527.05$ MeV.

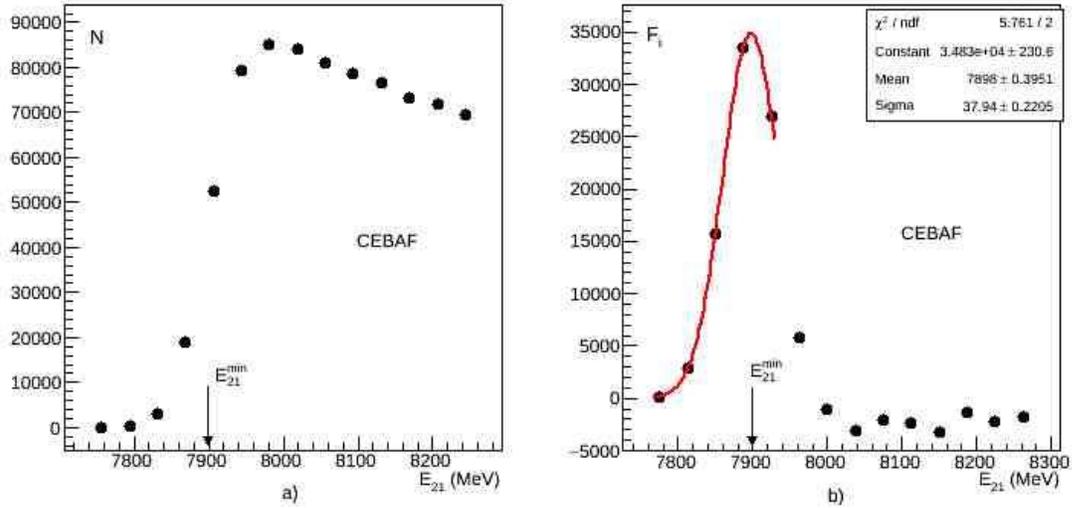

Fig. 8. a) MC energy spectrum of scattered electrons near Compton edge for Jlab experiment: $E_1 = 11000.0$ MeV, $\omega_{01} = 2.33$ eV. Total number of events is $5 \times 10^6$. b) Distribution of the $Y_i = N_{i+1} - N_i$ values. The extracted electron Compton edge is $E_{21}^{min} = 7898 \mp 0.4$ MeV with an expected theoretical $E_{21}^{min} = 7898.80$ MeV.

In order to obtain the absolute value of the Compton Edge we need to know absolute value of magnetic field integral in D3. It can be determined with accuracy $10^{-3} - 10^{-4}$ [38].

The absolute value of the energy of incident electrons can be determined also by means of the determination of the rations: 1. $y_e^{min} = E_{21}^{min}/E_1$; 2. $y_\gamma^{max} = \omega_{21}^{max}/E_1$; 3. $a_e = E_{21}^{min}/E_{22}^{min}$; 4. $a_\gamma = \omega_{22}^{max}/\omega_{21}^{max}$, where $E_{21}^{min}$, $E_{22}^{min}$, ($\omega_{21}^{max}$, $\omega_{22}^{max}$) are the values of Compton Edges of scattered electrons (photons) from two different laser lines ($\omega_{01}, \omega_{02}$) and $y_e = 1 - y_\gamma$; $y_e^{min} = 1 - y_\gamma^{max} = 1/(1+x)$ [33].

For example in the case of 2 and 3 we have

$$x = \frac{y_\gamma^{max}}{1 - y_\gamma^{max}} \tag{34}$$

and

$$x = \frac{1 - a_e}{a_e(1 - \omega_{02}/\omega_{01})}. \tag{35}$$

Therefore

## 5 GRAAL/ESRF: the Experimental Setup and measurements

The measurement of the Compton Edge of the scattered high energy electrons of synchrotrons on monochromatic laser beams, has been originally suggested in [14] as an efficient test for the one-way light speed isotropy and the Lorentz invariance in the reference frame of the cosmic microwave background radiation. This method has been successfully elaborated at GRAAL facility at the European Synchrotron Radiation Facility. Initially, the analysis of the data of 1998-2002 (non-continuous) measurements enabled to obtain an upper limit for the anisotropy $10^{-12}$ [15]. Then, dedicated measurements, i.e. with a facility upgraded for that particular goal, have been performed in 2008, which enabled to lower further that limit, up to $10^{-14}$ [16, 17].

In the experiment carried out with the GRAAL facility, installed at the European Synchrotron Radiation Facility (ESRF), the *γ*-ray beam was produced by Compton scattering of laser photons off the 6.03 GeV electrons circulating in the storage ring. Incoming photons are generated by a high-power Ar laser located about 40 m from the intersection region. The laser beam enters the vacuum via an MgF window and is then reflected by an Al-coated Be mirror towards the electron beam. The laser and electron beams overlap over a 6.5 m long straight section. Photons are finally absorbed in a four-quadrant calorimeter, which allows the stabilization of the laser-beam center to 0.1 mm. This level of stability is necessary and corresponds to a major improvement of the updated set-up. Because of their energy loss, scattered electrons are extracted from the main beam in the magnetic dipole following the straight section. Their position can then be accurately measured in the tagging system (Fig. 11) located 50 cm after the exit of the dipole.

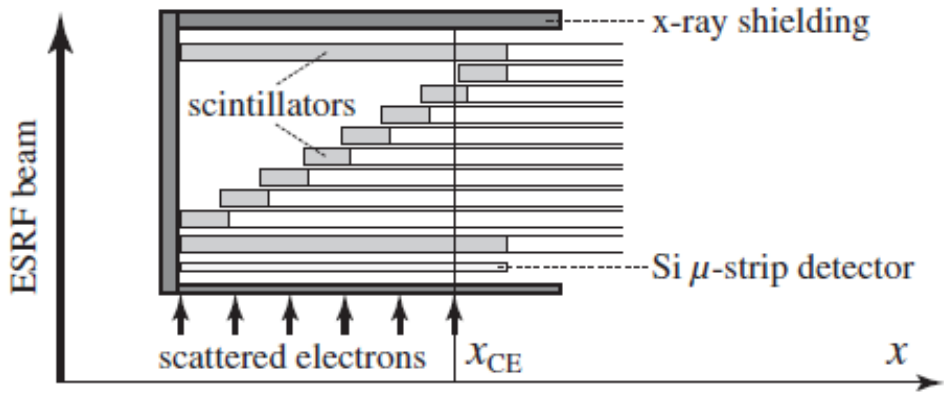

Fig. 9. The schematic of the GRAAL tagging system.

This system plays the role of a magnetic spectrometer from which we can infer the electron momentum. The tagging system is composed of a position-sensitive Si μ-strip detector (128 strips of 300 μm pitch, 500 μm thick) associated to a set of fast plastic scintillators for timing information and triggering of the data acquisition. These detectors are placed inside a movable box shielded against the huge x-ray background generated in the dipole. The x-ray induced heat load, which is the origin of sizable variations in the box temperature, correlated with the ESRF beam intensity. This produces a continuous drift of the detector due to the dilation of the box. A typical Si μ-strip count spectrum near the CE is shown in Fig. 12 for the green and multiline UV mode of the laser used in this measurement.

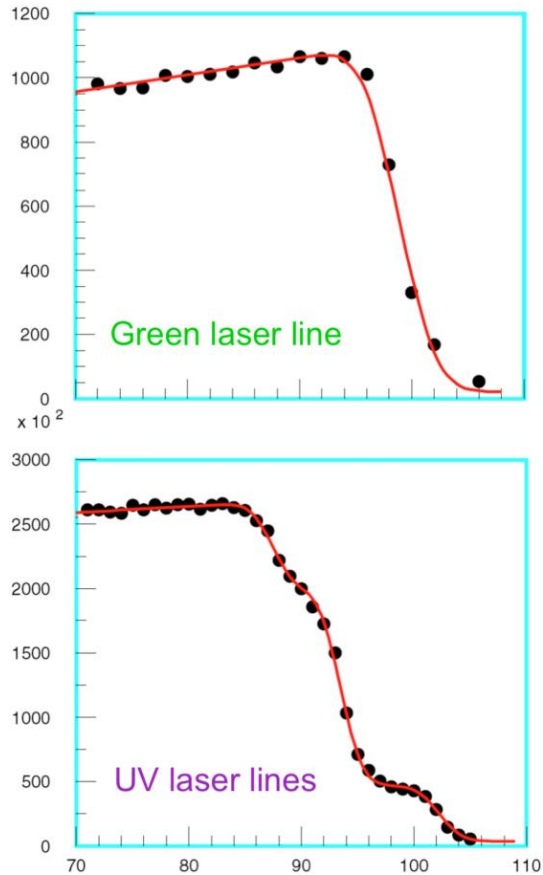

Fig. 10. The Compton Edge for the Green laser line (2.41 eV) and the lower one with three UV lines around 3.53 eV for the parameters of the GRAAL facility. The abscissa indicates the microstrip number.

The multiline UV mode (displayed separately in Fig. 13) corresponds to three groups of lines centered around 364, 351, and 333 nm, which are clearly resolved. The fitting function is also plotted in Fig. 13. The CE position, $x_{CE}$, is taken as the location of the central line. The steep slope of the CE permits an excellent measurement of $x_{CE}$ with a resolution of ~3 μm for a statistics of about $10^6$ counts.

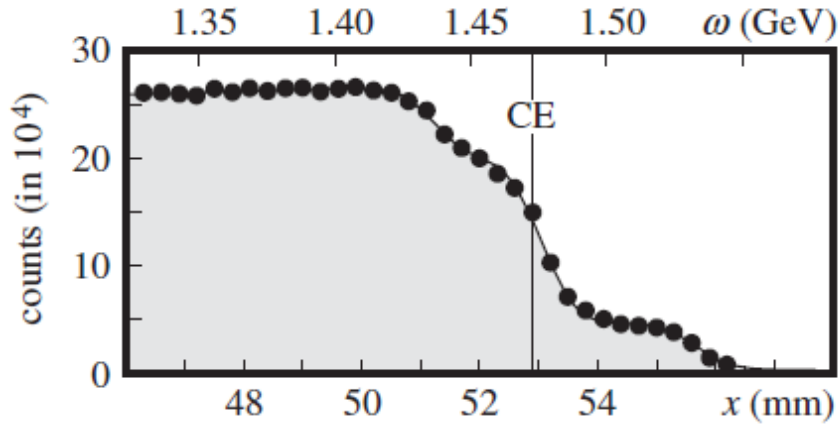

Fig. 11. Si μ-strip count spectrum near the CE and the fitting function vs position x and scattered photon energy ω.

During a week of data-taking in July 2008, a total of 14 765 CE spectra have been recorded. A sample of the time series of the CE positions relative to the ESRF beam covering 24 h is displayed in Fig. 14. Fig. 14(c), along with the tagging-box temperature [Fig. 14(b)] and the ESRF beam intensity [Fig. 14(a)]. The sharp steps present in Fig. 14(a) correspond to the twice-a-day refills of the ESRF ring. The similarity of the temperature and CE spectra combined with their correlation with the ESRF beam intensity led to interpret the continuous and slow drift of the CE positions as a result of the tagging-box dilation. To remove this trivial time dependence, a special fitting procedure was developed. The corrected and final spectrum, obtained by subtraction of the fitted function from the raw data, is plotted in Fig. 14(d). A sample of such a spectrum is displayed separately in Fig. 15.

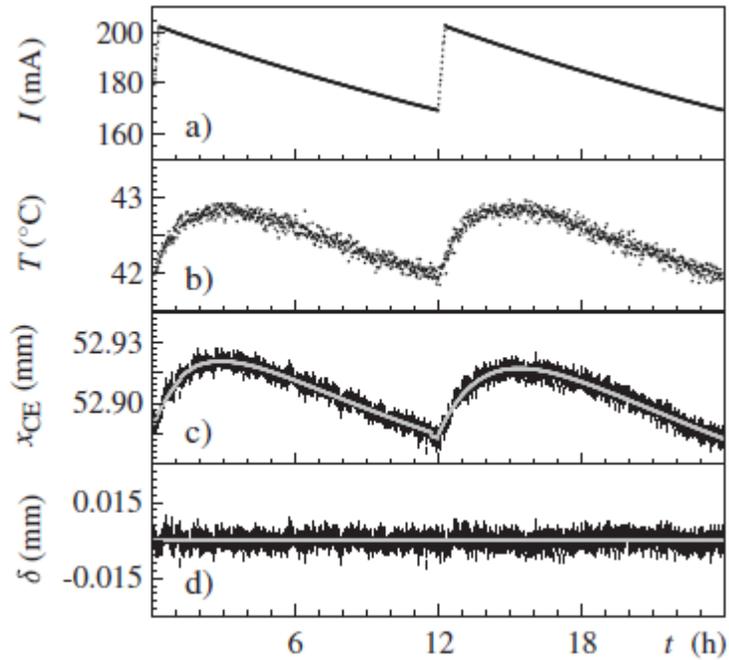

Fig. 12. Time evolution over a day of (a) ESRF beam intensity, (b) tagging-box temperature, (c) CE position and fitted curve, and (d) $\delta = x_{CE} - x_{fit}$. The error bars on position measurements are directly given by the CE fit.

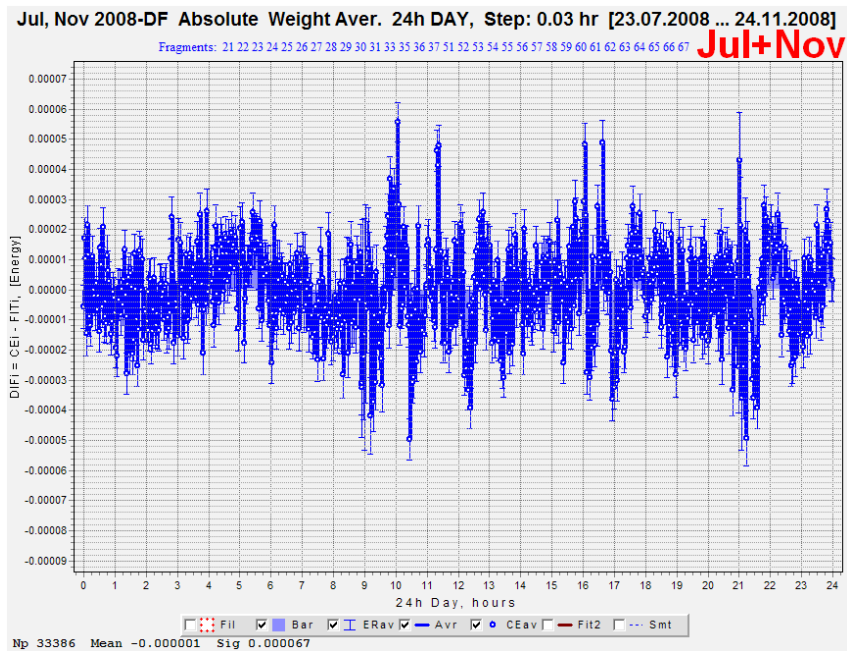

Fig. 13. The Compton Edge time variations obtained at GRAAL 2008 measurements.

The GRAAL collaboration presented an upper bound on a hypothetical sidereal oscillation of the CE energy to be less than $2.5 \cdot 10^{-6}$ (95% C. L.) yielding the competitive limit on the one way light speed anisotropy to be less than $1.8 \cdot 10^{-14}$ (95% C. L.).

## 6 Prospects for Lorentz Invariance tests with JLab (Møller) Compton polarimeter

The JLab Hall A/C Compton polarimeter is ideally suited for an advanced Lorentz Invariance tests by means of Laser Compton Backscattering. The Hall A/C Compton polarimeter setup is schematically presented in Fig. 14.

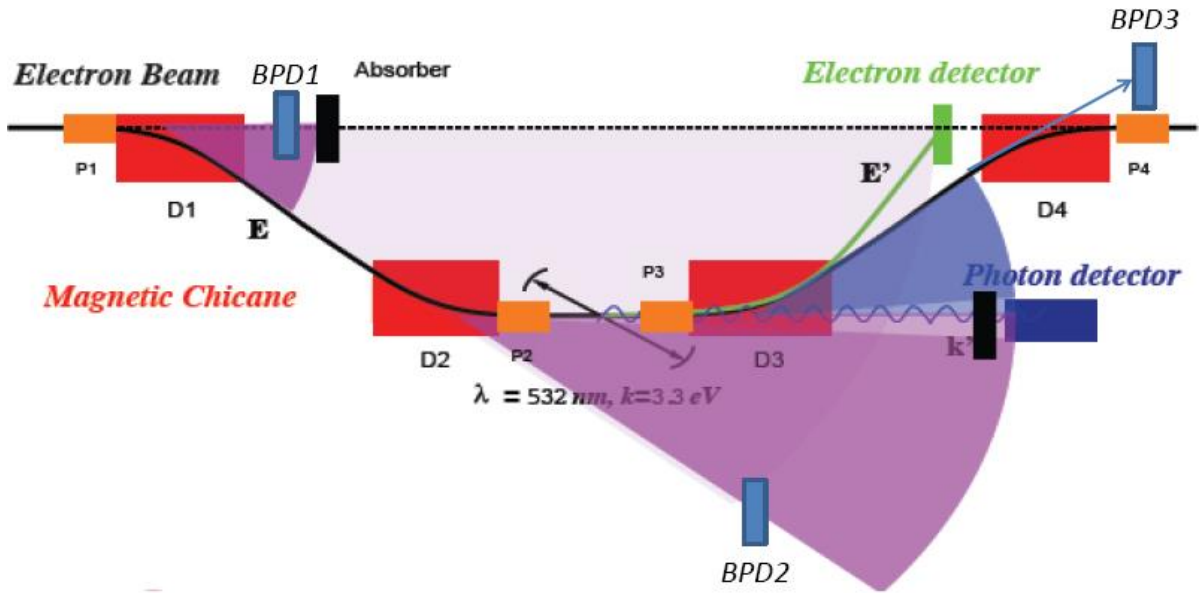

Fig. 14. Schematic layout of the Compton polarimeter and illustration of synchrotron radiation suppression scheme with fringe field modifying field plates P1-P4, attached to dipole magnets D1-D4 and beam position detectors BPD1-BPD3 (the main figure is depicted from Hall A Compton polarimeter- OSP, 2013 [39]).

**Compton Polarimeter**. The polarimeter provides electron beam polarization measurements in continuous and non-intrusive manner using Compton scattering of polarized electrons from polarized photons.

The primary features of the Compton polarimeter are:
1. A vertical magnetic chicane with four dipole magnets to transport the CEBAF electron beam to the Compton Interaction Point (CIP).

2. A high-finesse Fabry-Perot (FP) cavity serving as the photon target, located at the lower straight section of the chicane with the cavity axis at an angle of 24 mr with respect to the electron beam.
3. An electromagnetic calorimeter to detect the back-scattered photons.
4. A Silicon micro-strip electron detector to detect the recoil electrons, dispersed from the primary beam by the third dipole of the chicane.

The electron beam polarization is deduced from the counting rate asymmetries of the detected particles. The electron and the photon arms provide redundant measurement of the electron beam polarization.

As part of the 12 GeV upgrade of CEBAF, the Hall A Compton polarimeter has been reconfigured to accommodate the 11 GeV electron beam available to Hall A. The primary changes to the Compton polarimeter for the 12 GeV upgrade are:

a) Reduction of the Chicane displacement from 300 to 215 mm. The change in geometry allows the 11 GeV beam to be transported through the existing dipole magnets while necessitating the raising of the two middle dipole magnets, the optics table, and the photon detector by 85 mm.
b) Increase in the electron arm acceptance to allow detection of Compton edge in the electron detector with green laser photons.
c) Synchrotron radiation blocker for the electron detector in the straight through beam line after the first chicane dipole.
d) Suppression of synchrotron radiation background for the photon calorimeter with addition of field plates P1-P4 (see Fig. 14) to all four dipole magnets that soften the fringe fields seen by the photon detector.

The green FP cavity power has been boosted to over 10 kW with new low loss mirrors. The increased luminosity will provide head-room for high signal-to-background ratio for Compton scattering even with the anticipated higher beam background at 11 GeV, arising from larger momentum spread of the electron beam due to synchrotron radiation losses. The backscattered photons are transported to the position sensitive electromagnetic calorimeter via a telescoping beam pipe with a maximum diameter of 1.5 inch. The beam pipe is terminated with a vacuum window and a lead collimator with configurable absorbers to stop soft photons including synchrotron radiation. This configuration provides adequate acceptance from 1 to 11 GeV. As shown in Fig. 14, the Compton polarimeter consists of four major subsystems. The subsystems of the Compton Polarimeter are described below.

**Magnetic Chican**e. The Compton magnetic chicane, illustrated in Fig. 14, consists of 4 dipoles (1.5 T maximum field, 1 meter magnetic length) here after called D1,2,3,4. (D1, D2) deflect the electrons vertically down to steer the beam through the Compton interaction point (CIP) located at the center of the optical cavity. After the CIP, the electrons are vertically up deflected (D3,D4) to reach the Hall A target. The scattered electrons are momentum analyzed by the third dipole and detected thanks to 4 planes of silicon strips. The magnetic field is scaled with the beam energy, insuring the same vertical deflection at the CIP, up to 11 GeV electrons for 1.5 T field. The parameters of the Chicane are as follows:

- The longitudinal magnetic length on the axis of D1 and D2 is 1000 mm.
- The distance between the geometrical axis of the dipoles D1 and D2 in the longitudinal plane is 5400 mm.
- The distance between the beam entry axis in D1 and the beam exit axis in D2 in the bending plane (vertical axis), also known as the chicane displacement, is 215 mm.
- The bending angle is 2.35 degree.

With higher energy of the 12 GeV Upgrade, synchrotron radiation in the Compton chicane increases dramatically both in flux and energy leading to dilution of the Compton scattering signal in the detectors. The synchrotron radiation can be suppressed with the addition of passive iron plates in the fringe field region of the dipole magnets to reduce the magnetic field seen by the detectors, thus reducing synchrotron radiation background to manageable level. Shown in Fig. 15 is a schematic representation of the synchrotron radiation background and its suppression scheme. Dipole magnet D1 poses a potential source of synchrotron radiation for the electron detector via the straight through beam line, where as D2 and D3 produce similar background for the photon detector. These radiations will be softened with the addition field plates and reduced in flux with absorbers. Dipole magnets D1-D4 have been modified with fringe field plate P1-P4. Lead and/or Iron absorbers, matched to the beam energy, are installed external to the scattered photon beam line, for the photon detector.

We propose to add Beam Position Monitors BPD1-3 (see Fig. 14) to measure and control the directions of the incident and bended electron beams.

**Photon Target**. A high-finesse Fabry-Perot cavity housed on a optics table serves the role of the photon target. The optical setup consists of four parts:

1. Green Laser operating at 532 nm wavelength generating up to 3 W power;
2. Input optical transport form the laser beam to the cavity to optimize laser beam size and polarization;
3. The resonant Fabry-Perot cavity that delivers more than 10 kW of green light, with the cavity axis at an angle of 24 mr with respect to the electron beam;
4. Optical devices to measure the circular polarization of the photons at the exit of the cavity.

Details can be found in [39].

**Photon Detector**. To detect the Compton backscattered photons, an electromagnetic calorimeter is used. The backscattered photons are transported to the calorimeter via a telescoping beam pipe with a maximum diameter of 1.5 inch. The beam pipe is terminated with a vacuum window and a lead collimator with configurable absorbers to stop soft photons including synchrotron radiation. This configuration provides adequate acceptance from 1 to 11 GeV. We propose to use position sensitive photon detector. Since the Compton scattered photon angles near the Compton Edge are smaller ($\leq 10^{-5}$ rad, see Fig. 4) than the angular divergence of the beam ($10^{-4} - 10^{-5}$), the scattered photons moves with the same direction as electrons. So by measurement of the positions of scattered photons near the Compton Edge, direction of the incident beam can be determined and monitored. The position of each photon can be determined with few mm precision. Consequently, the centroid for $10^6$ photons can be determined with precision of a few $\mu m$.

**Electron Detector**. The electron detector is made up of 4 planes of Silicon micro-strips composed of 192 strips each. The micro-strips have 240 μm pitch (200 μm Silicon, and 40 μm spacing), on a 500 μm thick Silicon substrate, manufactured by Canberra systems. The planes are staggered by 80 microns to allow for better resolution. Shown in Fig. 10 is a schematic view of the electron detector. The detector is mounted in a vacuum chamber on a vertically movable shaft. A motion control system moves the detector to the appropriate location for the detection of Compton scattered electrons for a given electron beam energy. The detector can be positioned as close as 4 mm to the primary electron beam in order to allow for low energy Compton polarimetery. Distance between the CIP and the first strip is 5750 mm. We recall that between the CIP and the end of the Dipole 3 is 2150 mm, i.e. distance between the center of the Dipole 3 and the first strip is 4100 mm. For a beam of 11000.0 MeV the Compton edge is at 7898.8 MeV. The bending angles of the 11000.0 MeV and 7898.8 MeV electrons are 2.35° and 3.27° respectively. The single strip of the electron detector covered angular interval $\Delta\alpha = \frac{0.2}{4100} = 0.0028°$, or energy interval $\Delta E \cong 0.0028° \times \frac{7898.8}{3.27°} = 6.7$ MeV. The energy spread of the Compton scattered electrons are determined by energy and angular divergences of the incident electron and laser beams. These effects does not change the maximum energy, i.e. the Compton Edge of the LCS γ-photons and can be effectively incorporated into their energy spread or taken into account in MC code. As follows from the Monte Carlo simulations, the absolute value of mean energy of the beam can be determined within an order of magnitude better than the energy spread of the recoil electrons. At the GRAAL the Compton Edge was determined with in error of about $3\mu$ for a statistics of $10^6$ counts. At JLAB such a resolution can be achieved with less statistics, since the energy resolution of the tagging system at JLAB is expected to be better.

**Beam Position Detectors**. Deflection of the beam in a magnetic field is the simplest method for a momentum measurement. Such a measurement could provide very accurate monitoring of slight changes of momentum assuming stability of the magnetic field and the beam position detectors. Therefore the electron beam momentum can be measured and controlled with relative

We consider the case of measuring of the beam momentum and momentum distribution of recoil electrons near the Compton Edge. Basically the method involves measuring the deflection of recoil electrons in a magnetic field. Therefore, this method requires measurement of the magnetic field integrals and the bending angles. In order to determine the absolute value of the beam momentum and Compton Edge to $10^{-4}$, both of these quantities must be determined at a level somewhat better than $10^{-4}$. Three beam position detectors BPD1-BPD3 are proposed to use (see Fig.14). By using these detectors $10^{-6}$ relative precision over a period of one millisecond can be achieved [40]. The GRAAL experiment and our MC simulations demonstrated that by momentum measurement of recoil electrons the Compton edge can be determined again with $10^{-6}$ relative precision.

Currently, the attainable accuracy for the absolute value of the field integrals is of the order of $10^{-3}$-$10^{-4}$.

Since the Compton electron scattering angles are smaller ($\leq 10^{-5}$ rad, see Fig. 4) than the angular divergence of the beam ($10^{-4} - 10^{-5}$), the scattered and unscattered electrons remain unsepareted until they pass through D3 dipole magnet. Both electrons are dispersed and recoil electrons detected by electron detector (see Fig. 14).

In this manner with a modified polarimeter, main amount of data will be obtained working in parasitic mode, i.e. in this stage without requiring a dedicated beam time, and as show the simulations, we expect to improve about an order of magnitude the upper bound of the one way light speed anisotropy obtained at GRAAL.

Conclusions

The main goals expected for this project can be summarized as follows:
1. Based on successfully performed GRAAL-ESRF measurements and the obtained results, the Compton Edge method has proved its efficiency in testing the fundamental physical principles, i.e. one-way light speed isotropy and the Lorentz invariance violation.
2. By means of Compton Edge studies on 12 GeV electron accelerator beam in Jefferson National Laboratory, by **conservative** evaluation the simulations' results, the expected accuracy increase is about **an order of magnitude for the one-way light speed anisotropy** with respect the limit reached at GRAAL, as well as for the light speed's sensitivity for the velocity of apparatus (Kennedy-Thorndike experiment) of accuracy higher than the available limits or those proposed for satellite experiments.
3. The expected results of higher accuracy of the Compton Edge relative changes will have a direct impact on theoretical models of Lorentz invariance violation, will enable either to close certain models or obtain constraints on the parameters of the others. In outcome, the models will affect numerous fundamental physical problems including in cosmology and the evolution of the very early Universe.

The ongoing studies of the cosmic microwave background by ground based telescopes and Planck satellite and the forthcoming data for tracing up to the sub-Planckian scale of energies with the B-mode polarization, non-Gaussianities [41-45], dark energy redshift evolution (e.g. [46]), blazar spectra [47], satellite tests of General Relativity [48-50], make more outlined this proposal, along with the traditional interest for Lorentz invariance violating models. The recent discovery of the gravitational waves [51] also has been associated to the Lorentz violating models and their tests [52].

In the present Proposal, as compared to the LoI submitted a year ago, more explicit illustrations of the observables are obtained at simulations for the input parameters available at JLab. However, all those are no more than illustrations, and the key basis to ensure both, a successful experiment and data analysis, is our GRAAL-ESRF experience, which had led to results acting currently as reference numbers for theoretical models. The next point to be stressed is that, various tests of light speed isotropy and Lorentz invariance violations have to be considered not as mutually competing ones, but, on the contrary, as complementing each other reflecting the involved physical processes, associated in relevant (not same!) ways to those fundamental physical principles.

To achieve the goal this Proposal is stating the need of: (a) electron and photon detectors ensuring the needed accuracy of the measurements, and (b) the beam time. Both in view of the experience with GRAAL-ESRF, look absolutely feasible in JLab.

Regarding practical issues two options are possible:

1. use of the available detectors;
2. development of new detectors compatible with the precision polarization measurements required, which, certainly, will need some efforts.

We intend to start with the first option by using the existing experimental setup at HALL A/C and without requesting a dedicated beam time for the data storage.

**References**

1. A. Einstein, NeueExperimenteüber den Einfluβ der Erdbewegung and die LichtgeschwindigkeitrelativzurErde, Forsch. Und Fortschritte, 3, 36, 1927.
2. R. F. C. Vessot, M. W. Levine, E. M. Mattison, E. L. Blomberg, T. E. Hoffman, G. U. Nystrom, B. F. Farrel, R. Decher, P. B. Eby, C. R. Baugher, J. W. Watts, D. L. Teuber, and F. D. Wills, Test of Relativistic Gravitation with a Space-Borne Hydrogen Maser, Phys.Rev.Lett. 45, 2081, 1980.
3. E. Riis, L-U. A. Andersen, N. Bjerre, O. Poulsen, S. A. Lee, and J. L. Hall, Test of the Isotropy of the Speed of Light Using Fast-Beam Laser Spectroscopy, Phys.Rev.Lett. 60, 81, 1988.
4. Z. Bay, J. White, Comment on Test of the Isotropy of the Speed of Light, Phys.Rev.Lett. 62, 841, 1989.
5. S. Herrmann , A. Senger, E. Kovalchuk, H. Müller, A.Petersetal, Test of the Isotropy of the Speed of Light Using a Continuously Rotating Optical Resonator, Phys. Rev. Lett. 95,150401, 2005.
6. P. Antonini, M. Okhapkin, E. Göklü, and S. Schiller, Test of constancy of speed of light with rotating cryogenic optical resonators, Physical ReviewA71, 050101, 2005.
7. M. A. Hohensee, P. L. Stanwix, M. E. Tobar, S. R. Parker, D. F. Phillips, R. L. Walsworth, Improved constraints on isotropic shift and anisotropies of the speed of light using rotating cryogenic sapphire oscillators", Physical Review D**82**, 076001, 2010.
8. V.A. Kostelecky, Gravity, Lorentz violation, and the standard model, Phys. Rev. D. 2004. – 69, 105009, 2004.
9. D. Mattingly, Modern Tests of Lorentz Invariance, Living Rev. Relativity,8, 5, 2005.
10. V.A. Kostelecky, N. Russell, Data Tables for Lorentz and CPT Violation, Reviews of Modern Physics,83, 2011.
11. C. Lammerzahl et al, Kinematical test theories for Special Relativity: A comparison, Int. J. Mod. Phys. D., 11, 1109, 2002.
12. BICEP2 Collaboration: P.A.R. Ade, R. W. Aikin, D. Barkats, S. J. Benton, C. A. Bischoff, J. J. Bock, J. A. Brevik, I. Buder, E. Bullock, C. D. Dowell, L. Duband, J. P. Filippini, S. Fliescher, S. R. Golwala, M. Halpern, M. Hasselfield, S. R. Hildebrandt, G. C. Hilton, V.

V. Hristov, K. D. Irwin, K. S. Karkare, J. P. Kaufman, B. G. Keating, S. A. Kernasovskiy, J. M. Kovac, C. L. Kuo, E. M. Leitch, M. Lueker, P. Mason, C. B. Netterfield, H. T. Nguyen, R. O'Brient, R. W. Ogburn IV, A. Orlando, C. Pryke, C. D. Reintsema, S. Richter, R. Schwarz, C. D. Sheehy, Z. K. Staniszewski, R. V. Sudiwala, BICEP2 I: Detection Of B-mode Polarization at Degree Angular Scales, arXiv:1403.3985

13. C. R. Contaldi, M. Peloso, L. Sorbo, Suppressing the impact of a high tensor-to-scalar ratio on the temperature anisotropies, arXiv:1403.4596

14. V.G. Gurzadyan, A.T. Margarian, Inverse Compton testing of fundamental physics and the cosmic background radiation, Physica Scripta, 53, 513, 1996.

15. V.G. Gurzadyan, J.P. Bocquet, A.Kashin, A.Margarian, O.Bartalini, V.Bellini, M.Castoldi, A.D'Angelo, J.-P.Didelez, R.Di Salvo, A.Fantini, G.Gervino, F.Ghio, B.Girolami, A.Giusa, M.Guidal, E.Hourany, S.Knyazyan, V.Kouznetsov, R.Kunne, A.Lapik, P.LeviSandri, A.Lleres, S.Mehrabyan, D.Moricciani, V.Nedorezov, C.Perrin, D.Rebreyend, G.Russo, N.Rudnev, C.Schaerf, M.-L.Sperduto, M.-C.Sutera, A.Turinge, Probing the Light Speed Anisotropy with respect to the Cosmic Microwave Background Radiation Dipole, Mod. Phys. Lett. A, 20,1, 2005.

16. V.G. Gurzadyan, J.P.Bocquet, A.Kashin, A.Margarian, O.Bartalini, V.Bellini, M.Castoldi, A.D'Angelo, J.-P.Didelez, R.Di Salvo, A.Fantini, G.Gervino, F.Ghio, B.Girolami, A.Giusa, M.Guidal, E.Hourany, S.Knyazyan, V.Kouznetsov, R.Kunne, A.Lapik, P.LeviSandri, A.Lleres, S.Mehrabyan, D.Moricciani, V.Nedorezov, C.Perrin, D.Rebreyend, G.Russo, N.Rudnev, C.Schaerf, M.-L.Sperduto, M.-C.Sutera, A.Turinge, Lowering the Light Speed Isotropy Limit: European Synchrotron Radiation Facility Measurements, NuovoCimento, 122, 515, 2007.

17. J.-P. Bocquet, D. Moricciani, V. Bellini, M. Beretta, L. Casano, A. D'Angelo, R. Di Salvo, A. Fantini, D. Franco, G. Gervino, F. Ghio, G. Giardina, B. Girolami, A. Giusa, V.G. Gurzadyan, A. Kashin, S. Knyazyan, A. Lapik, R. Lehnert, P. Levi Sandri, A. Lleres, F. Mammoliti, G. Mandaglio, M. Manganaro, A. Margarian, S. Mehrabyan, R. Messi, V. Nedorezov, C. Perrin, C. Randieri, D. Rebreyend, N. Rudnev, G. Russo, C. Schaerf, M.L. Sperduto, M.C. Sutera, A. Turinge, V. Vegna, Limits on light-speed anisotropies from Compton scattering of high-energy electrons, Phys.Rev.Lett.104:241601, 2010.

18. V.G. Gurzadyan, V. Bellini, M. Beretta, J.-P. Bocquet, A. D'Angelo, R. Di Salvo, A. Fantini, D. Franco, G. Gervino, G. Giardina, F. Ghio, B. Girolami, A. Giusa, A. Kashin,

H.G. Khachatryan, S. Knyazyan, A. Lapik, P. Levi Sandri, A. Lleres, F. Mammoliti, G. Mandaglio, M. Manganaro, A. Margarian, S. Mehrabyan, R. Messi, D. Moricciani, V. Nedorezov, D. Rebreyend, G. Russo, N. Rudnev, C. Schaerf, M.-L. Sperduto, M.-C. Sutera, A. Turinge, V. Vegna, A new limit on the light speed isotropy from the GRAAL experiment at the ESRF, Proceed. XII M. Grossmann meeting on General Relativity, World. Sci., vol. B, p.1495, 2012.; arXiv:1004.2867

19. R.J. Kennedy, E.M. Thorndike, Experimental Establishment of the Relativity of Time, Phys. Rev. **42**, 400, 1932.

20. J. A. Lipa, S. Buchman, S. Saraf, J. Zhou, A. Alfauwaz, J. Conklin, G. D. Cutler and R. L. Byer, Prospects for an advanced Kennedy-Thorndike experiment in low Earth orbit, arXiv: 1203.3914.

21. M. E. Tobar, P.Wolf, S. Bize, G. Santarelli, and V. Flambaum, Testing local Lorentz and position invariance and variation of fundamental constants by searching the derivative of the comparison frequency between a cryogenic sapphire oscillator and hydrogen maser, Phys. Rev. D, **81**, 22003 (2010).

22. C.L. Bennett et al, Nine-Year Wilkinson Microwave Anisotropy Probe (WMAP) Observations: Final Maps and Results, ApJS, 208, 20, 2013

23. S. Rauzy, V.G. Gurzadyan, On the motion of the Local Group and its substructures, Mon. Not. Roy. Astr. Soc., 298, 114, 1998.

24. M.F. Ahmed, B.M. Quine, A. Sargoytchev, S.Stauffer, A. D.,A review of one-way and two-way experiments to test the isotropy of the speed of light, I. Journ. Phys., 86, 835, 2012.

25. L. Lingli, B. Ma, A theoretical diagnosis on light speed anisotropy from GRAAL experiment, Astroparticle Phys., 36, 37, 2012.

26. M. Cambiaso, R. Lehnert, R. Potting, Massive Photons and Lorentz Violation, Phys. Rev. D, 85, 085023, 2012.

27. V.A. Kostelecky, D. Mewes, Electrodynamics with Lorentz-violating operators of arbitrary dimension, Phys. Rev. D. 80, 015020, 2009.

28. B. Audren, D. Blas, J. Lesgourgues, S. Sibiryakov, Cosmological constraints on Lorentz violating dark energy, Journal of Cosmology and Astroparticle Physics, 08, id. 039, 2013.

29. B. Audren, D. Blas; M. M. Ivanov, J. Lesgourgues, S. Sibiryakov, Cosmological constraints on deviations from Lorentz invariance in gravity and dark matter, arXiv:1410.6514, 2014.
30. S. Ando, M. Kamionkowski, and I. Mocioiu, Neutrino Oscillations, Lorentz/CPT Violation, and Dark Energy, Phys.Rev. D80, 123522,2009
31. C. Zhe, S. Wang, Constraints on space time anisotropy and Lorentz violation from the GRAAL experiment, Eur. Phys. J. C, 73, 2337, 2013.
32. I. Ginzburg et al. Colliding gamma electron and gamma gamma Beams Based on the Single Pass Accelerators (of Vlepp Type)**,** Nucl. Instr. Meth. 205, 47, 1983.
33. S. Knyazian, A. Margarian, S. Mehrabyan. Compton Edge electron beam spectrometer, Preprint YerPhi 1486(3)-97, 1997.
34. Ch.W. Leemann, D.R. Douglas, and G.A. Kraft, Annual Review of Nuclear and Particle Science, 51, 413 (2001); CEBAF, http://www.scholarpedia.org/article/The_Thomas_Jefferson_National_Accelerator_Faciliy
35. T. P. Welch and R. Ent, Energy measurement of electron beams by Compton scattering, CEBAF, unpublished (1994).
36. W. R. Nelson, H. Hirayama, and W. O. Roger, "The EGS4 code systems," SLAC National Accelerator Lab., Menlo Park, CA, USA, SLAC-Report-265, 1985.
37. Robert Christophe, Mesure de la stabilite en energie du faisceau d'electrons de l'E.S.R.F. Rapport de stage, annee 1997-1998.
38. S. Watson et al., Precision measurements of the SLC reference magnets. SLAC-PUB-4908 (1989).
39. D. Gaskell, High Precision Compton and Møller Polarimetry at Jefferson Lab, talk at PAVI14, July, 2014.
40. P. Chvetsov, A. Day, J.-C. Denard, A. P. Freyberger, R. Hiks. Non-invasive energy spread monitoring for the JLAB experimental program via synchrotron light interferometers, Nucl. Instr. & Meth. A557, pp 324-327 (2006).
41. F. R. Bouchet, The Planck mission, arXiv:1405.0439
42. Planck Collaboration: P. A. R. Ade, N. Aghanim, M. Arnaud, M. Ashdown, J. Aumont, C. Baccigalupi, et al, Planck intermediate results. XVI. Profile likelihoods for cosmological parameters, arXiv:1311.1657

43. V.G. Gurzadyan, P.A.R. Ade, P. de Bernardis, C.L. Bianco, J.J. Bock, A. Boscaleri, B.P. Crill, G. De Troia, K. Ganga, M. Giacometti, E. Hivon, V.V. Hristov, A.L. Kashin, A.E. Lange, S. Masi, P.D. Mauskopf, T. Montroy, P. Natoli, C.B. Netterfield, E. Pascale, F. Piacentini, G. Polenta, J. Ruhl, Ellipticity analysis of the BOOMERanG CMB maps, Int. J. Mod. Phys. D12, 1859, 2003.
44. V.G. Gurzadyan, A.L. Kashin, H. Khachatryan, E. Poghosian, S. Sargsyan, G. Yegorian, To the center of cold spot with Planck, Astron. & Astrophys.566, id.A135, 2014.
45. V.G.Gurzadyan, A.A.Kocharyan, Porosity Criterion for Hyperbolic Voids and CMB, Astron. & Astrophys. 493, L61, 2009.
46. S.G. Djorgovski, V.G. Gurzadyan, Dark Energy From Vacuum Fluctuations, Nucl. Phys.B PS173, 6, 2007.
47. F. Tavecchio, G. Bonnoli, On the detectability of Lorentz invariance violation through anomalous multi-TeV γ-ray spectra of blazars, arXiv:1510.00980
48. I. Ciufolini, A. Paolozzi, E. Pavlis, R. Koenig, J. Ries, V. Gurzadyan, R. Matzner, R. Penrose, G. Sindoni, C. Paris, Preliminary orbital analysis of the LARES space experiment, Eur. Phys. J. Plus 130, 133, 2015.
49. I. Ciufolini, A. Paolozzi, E. C. Pavlis, R. Koenig, J. Ries, V. Gurzadyan, R. Matzner, R. Penrose, G. Sindoni, C. Paris, H. Khachatryan, S. Mirzoyan, A Test of General Relativity Using the LARES and LAGEOS Satellites and a GRACE Earth's Gravity Model, Eur. Phys. J. C, 76, 120, 2016.
50. V. G. Gurzadyan, I. Ciufolini, A. Paolozzi, A. L. Kashin, H. G. Khachatryan, S. Mirzoyan and G. Sindoni, Satellites testing general relativity: Residuals versus perturbations, Int. J. Modern Physics D, 26, 1741020, 2017.
51. B.P. Abbott et al, LIGO Scientific Collaboration and Virgo Collaboration, Observation of Gravitational Waves from a Binary Black Hole Merger, Phys. Rev. Lett. 116, 061102, 2016.
52. A. Kostelecky, M. Mewes, Testing local Lorentz invariance with gravitational waves, Phys. Lett. B757, 510, 2016.